\documentclass[aps,prl,twocolumn,superscriptaddress,floatfix]{revtex4}
\usepackage{amsmath,amssymb,graphicx,graphics,color}
\usepackage{dcolumn}% Align table columns on decimal point
\usepackage{bm}% bold math
\usepackage{psfrag}

\newcommand{\bs}{\boldsymbol}

\begin{document}

\title{Exploring spin-orbital models 
with dipolar fermions in zig-zag optical lattices}
\author{G. Sun}
\affiliation{Institut f\"ur Theoretische Physik, Leibniz Universit\"at Hannover, 30167~Hannover, Germany}
\author{G. Jackeli}
\altaffiliation[]{Also at Andronikashvili Institute of Physics, 0177 
Tbilisi, Georgia.}
\affiliation{Max-Planck-Institut f\"ur Festk\"orperforschung,
Heisenbergstrasse 1, D-70569 Stuttgart, Germany}
\author{L. Santos}
\affiliation{Institut f\"ur Theoretische Physik, Leibniz Universit\"at Hannover, 30167~Hannover, Germany}
 \author {T. Vekua}
\affiliation{Institut f\"ur Theoretische Physik, Leibniz Universit\"at Hannover, 30167~Hannover, Germany}

\begin{abstract}
Ultra-cold dipolar spinor fermions in zig-zag type
optical lattices can mimic spin-orbital models relevant in solid-state systems, as transition-metal oxides 
with partially filled $d$-levels,
%pyroxene titanium and layered vanadium oxides, 
with the interesting advantage of reviving the quantum nature of 
orbital fluctuations. We discuss two different physical systems in which these models may be simulated, showing that  
the interplay between lattice geometry and spin-orbital quantum dynamics produces a wealth of novel quantum phases.
\end{abstract}

\maketitle

\date{\today}

%\pacs{05.30.Fk, 03.75.Ss, 03.75.Mn, 71.10.Fd}

% Introduction

Orbital degrees of freedom of electrons play an important role in the 
formation of various novel phases observed in transition-metal oxides 
with partially filled $d$-levels \cite{Tok00,Dag05}. In Mott insulators, 
they may enhance thermal as well as quantum fluctuations~\cite{Feiner97} 
and lead to spin/orbital liquid 
states~\cite{Khaliulin00,Vishwanath,Mila11,Jackeli11}, and to spontaneously
dimerized states without any long-range magnetic 
order~\cite{Jackeli07,Jackeli08}. In many solid-state 
systems, the orbital dynamics is often quenched,
due to the coupling of the orbitals to Jahn-Teller phonons, and 
the study of the quantum nature of orbitals demands systems with 
strong super-exchange coupling between spins and orbitals. 
However, in real materials, the coupling strengths  are fixed by nature, 
being very difficult to modify, thus limiting the experimental access 
to a potentially vast phase diagram.

Ultra-cold spinor gases in optical lattices open new fascinating perspectives 
for the analysis of the quantum nature of 
orbitals~\cite{Wu07,Hermele09,Gorshkov10}. The coupling constants in these 
systems can be easily controlled by modifying the lattice parameters 
and/or by means of Feshbach resonances~\cite{Chin2010}. 
Different lattice geometries, including frustrated lattices, 
like triangular~\cite{Struck2011} and Kagom\'e lattices~\cite{Jo2011}, 
may be created by combining different counter-propagating laser beams, 
and superlattice techniques. In addition, not only the physics in the 
lowest band but also that in higher bands may be controllably 
studied~\cite{Wirth2011}. Moreover, recent experiments on 
Chromium~\cite{Lahaye2007} and Dysprosium~\cite{Lu2011} atomic gases, 
and polar molecules~\cite{Ni2010}, are unveiling the exciting physics 
of ultra-cold dipolar gases, for which the dipole-dipole interactions 
%are expected to 
may
lead to exotic phases~\cite{Goral2002,He11}. 

% This Letter
In this Letter, we show that dipolar spinor Fermi gases in appropriate 
zig-zag lattice geometries allow for the quantum simulation of 
spin-orbital models for a family of Mott insulating compounds, 
including systems with weak~\cite{Jackeli07,Konstantinovic04,Hikihara} 
as well as pronounced~\cite{Jackeli09} relativistic spin-orbit couplings.
Moreover, these models, which are relevant for real materials, as pyroxene
titanium and layered vanadium oxides, may be explored with dipolar Fermi gases 
in parameter regimes which are hardly accessible for solid-state compounds, 
allowing for the observation of novel quantum phases.

%%%%%%%%%%%%%%%%%%%%%%%%%%%%%%%

%% FIGURE 1

\begin{figure}%[ht]
\includegraphics[width=8.6cm]{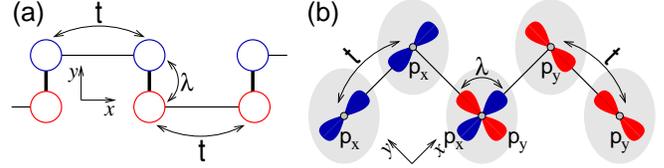}
\caption{(Color online) Two different systems discussed in this work. 
(a) Snake-like ladder with rungs (bonds) along $y$~($x$), and one fermion per rung. 
The inter (intra)-rung hopping is indicated by $t~(\lambda)$;
(b) Zig-zag lattice with one fermion per site occupying either $p_x$ or $p_y$ 
orbital states. Along a bond in $x$ $(y)$ direction only 
$p_x$ ($p_y$) orbitals are connected by a finite hopping amplitude $t$. 
A deformation of the lattice wells (shaded area) leads to a mixing between
$p_{x}$ and $p_{y}$ orbitals with amplitude $\lambda$.}
\label{lattices}
\end{figure}

%%%%%%%%%%%%%%%%%%%%%%%%%%%%%%%

% REALIZATIONS AND MODEL

{\it Physical realizations and effective Hamiltonian}.-- 
We start by presenting two possible scenarios, $\mathsf{A}$ and $\mathsf{B}$, 
in which dipolar Fermi gases may allow for the simulation of 
the above-mentioned spin-orbital models.

$\mathsf{A}$: First scenario is provided by a 
snake-like ladder as that shown in Fig.~\ref{lattices}(a), formed by bonds 
and rungs along $x$ and $y$ directions, respectively. This lattice geometry 
may be achieved by using a combined blue-detuned ladder-like lattice and a 
properly aligned red-detuned zig-zag lattice~\cite{footnote-lattices}. 
The overimposed red-detuned lattice allows for enhancing the potential barrier 
at alternated bonds in the upper and lower legs. A sufficiently strong
red-detuned lattice leads hence to broken bonds as depicted in
Fig.~\ref{lattices}(a). A similar technique has been recently employed for the 
realization of a Kagom\'e lattice~\cite{Jo2011}. The hopping within the same
rung is denoted by $\lambda$, whereas the inter-rung hopping for unblocked 
bonds is denoted as $t$ [Fig.~\ref{lattices}(a)]. The on-site repulsion $U$ 
results from the combination of dipolar and contact interactions. The dipoles 
are oriented in the $xz$-plane in such a way that fermions at the same rung 
experience maximal non-local repulsion $V$ and dipoles on neighboring rungs 
interact identically, irrespective whether they occupy upper or lower sites, 
with the interaction strength much weaker than intra-rung repulsion.
As a result, inter-rung dipolar interaction plays no role in the discussion 
below (interaction between next-nearest rungs is considered negligibly small). 
We assume $U,V \gg t, \lambda $, and consider the case of one fermion per
rung. The system is then in the Mott-insulator regime with one fermion
localized on each rung.
The sites belonging to the $i$-th rung of the ladder are distinguished by the
pseudo-orbital quantum number $\sigma_i^z$, with the convention that
$\sigma_i^z=+(-)1$ when the upper (lower) site on a given rung is occupied.
Defining $\alpha\equiv U/V$, and setting $t^2/2U$ as the energy unit 
we arrive to the effective spin-orbital Hamiltonian of Kugel-Khomskii 
type~\cite{KugelKhomskii,supplementary},
%%%%%%%%%%%%%
\begin{eqnarray}
\label{Themodel}
H&=\!&\sum_{i}^N[2 {\bf S}_i{\bf S}_{i+1}\!+\!\alpha-\!\frac{1}{2} ]
\left[1\!+\!(-1)^i \sigma_i^z][1\!+\!(-1)^i \sigma_{i+1}^z\right]\nonumber\\
&-&  \Delta \sum_{i}^N2{\bf S}_i{\bf S}_{i+1}
\left[1- \sigma_i^z \sigma_{i+1}^z\right]  -\lambda \sum_i^N \sigma_i^x,
\end{eqnarray}
%%%%%%%%%%%%%
where ${\bf S}_i$ is the spin-$\frac{1}{2}$ operator (stemming from the 
spinor nature of the Fermi gas), 
and $\sigma_i^{z,x}$ are Pauli matrices describing the pseudo-orbital
variables. We have added an additional 
term proportional to the coupling constant $\Delta$. Although $\Delta=0$ for
realization~$\mathsf{A}$, it plays an important role in the 
alternative case~$\mathsf{B}$ discussed below.
Note that the ratio $\alpha$ may be modified basically at will, since $U$ and
$V$ may be independently controlled using Feshbach resonances~\cite{Chin2010}, 
altering the lattice spacings, or modifying the transversal 
confinement~\cite{Goral2002}.

$\mathsf{B}$:  Second possible scenario is provided by a zig-zag lattice 
in the $xy$-plane, loaded with spinor dipolar fermions in 
$p$-bands~\cite{Wirth2011} [Fig.~\ref{lattices}(b)], where the dipoles 
are oriented along $z$-axis. Assuming a strong confinement along $z$, 
we retain two degenerate orthogonal $p_{x}$ and $p_{y}$ orbitals per lattice site. In this realization, $t$ denotes 
the hopping between similar orbitals at neighboring 
wells~[Fig.~\ref{lattices}(b)]. An in-plane deformation of the 
lattice wells (e.g. by an additional weak tilted lattice)
leads to a mixing of the $p_x$ and $p_y$ orbitals 
within the same well with an amplitude $\lambda$.

The interaction parameters for two fermions within
the same well are on-site repulsions $U$ (within the same orbitals) and $V$
(among different orbitals), and Hund's exchange $J_H$~\cite{supplementary}. 
Two fermions occupying the same orbital may form a symmetric or 
an antisymmetric 
state with respect to the orbital index with corresponding energies 
$U+J_H$ and $U-J_H$, which are split by the so-called pair-hopping term 
with an amplitude $J_H$~\cite{supplementary}. The overall energy scale will
now be modified to $t^2/2{\tilde U}$, where ${\tilde U}=(U^2-J_{H}^{2})/U$.
When two fermions occupy different orbitals, they may form a spin-singlet or 
a spin-triplet state with corresponding energies $V+J_H$ and $V-J_H$, 
which are split by Hund's exchange. In the strong coupling limit 
$U\pm J _H ,V\pm J_H \gg t, \lambda$  and with one particle per 
lattice well the system is in the Mott-insulator regime, and we arrive at
Hamiltonian~(\ref{Themodel}) with
$\Delta=J_H\tilde U/(V^2-J_{H}^{2})$~\cite{supplementary}. Now, $\alpha$ also  gets modified
accordingly, $\alpha= \tilde U(V+J_H/2)/(V^2-J_H^2)$.

Note that for a purely contact interaction $V=J_H$ and in the triplet channel 
two fermions do not experience any repulsion~\cite{supplementary}.
Thus, without dipolar couplings, the Mott phase of one fermion per 
well would not be stable. In the Mott-insulator regime, one can still 
vary $\Delta$ in a wide range by changing the relative ratio of the 
strengths of the contact and dipolar interactions.  

The model~\eqref{Themodel}, at $\alpha \simeq 1$,  describes the 
spin-orbital interplay in Mott insulating transition-metal compounds. 
At $\lambda=0$, it describes a 
zig-zag chain of spin one-half Ti$^{3+}$ ions, with active $d_{xy}$ and
$d_{yz}$  orbitals, in pyroxene titanium oxides 
$\mathrm{ {\it A}TiSi_2O_6\, ({\it A}=Na,Li)}$~\cite{Konstantinovic04,Hikihara}. 
On the other hand, for $\Delta=0$ the Hamiltonian~\eqref{Themodel} represents
the 1D counterpart of the 2D  model for $\mathrm {Sr_2VO_4}$~\cite{Jackeli09}. 
In the latter, the role of spins in Eq.~\eqref{Themodel} is played by an 
isospin variable discerning the Kramers partners, while the pseudo-orbital 
variables distinguishes two
lowest Kramers doublets of V$^{4+}$ ion, 
and $\lambda$ term in Eq.~\eqref{Themodel} represents relativistic 
spin-orbit coupling.

In the following we study the ground-state properties of the
model~\eqref{Themodel} in the wide parameter range by analytical 
and complementary numerical approaches.
%%%%%%%%%%%%%%%%%%%%%%%%%%%%%%%%%

\begin{figure}%[ht]
\includegraphics[width=4.25cm]{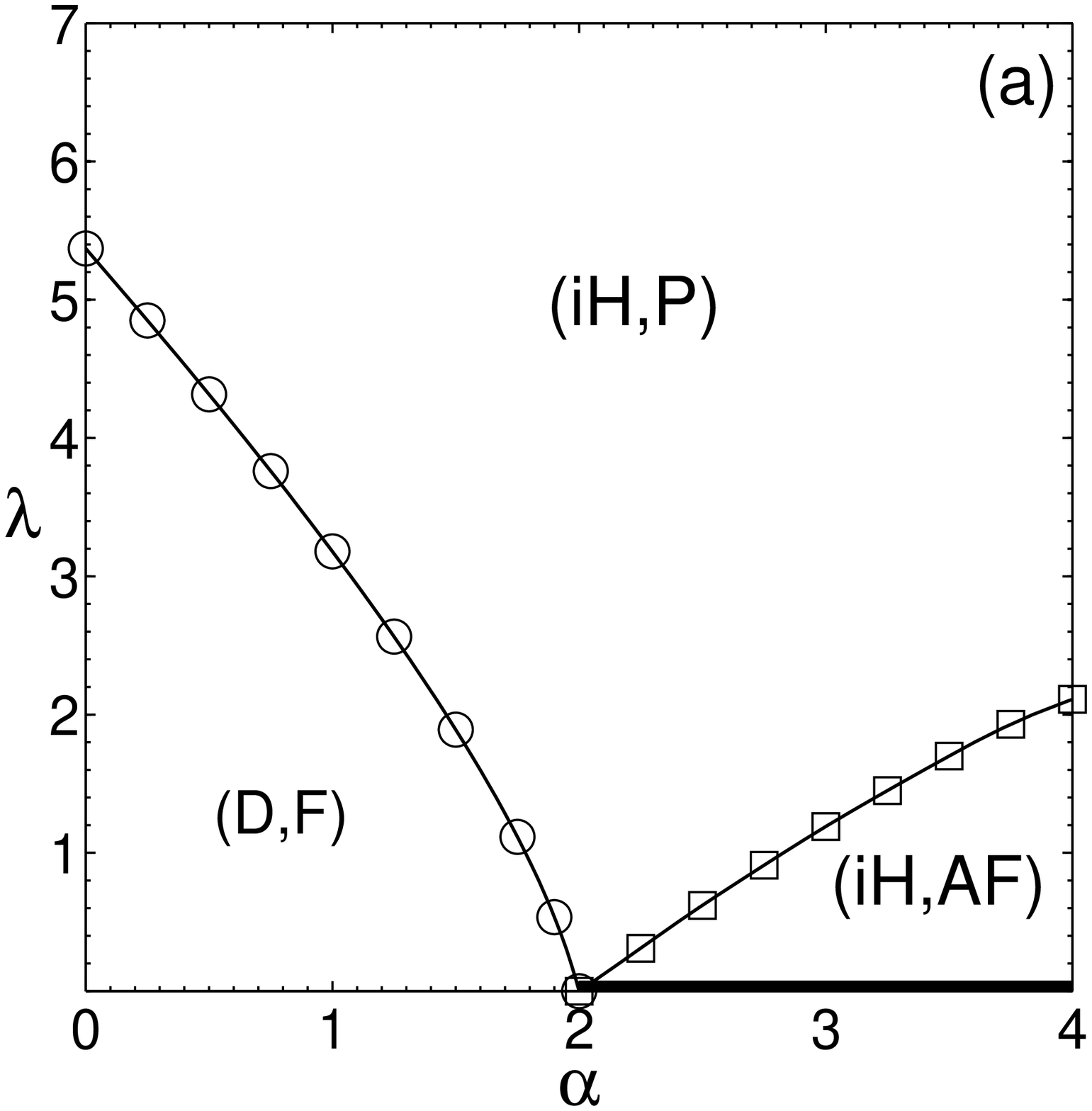}
\includegraphics[width=4.25cm]{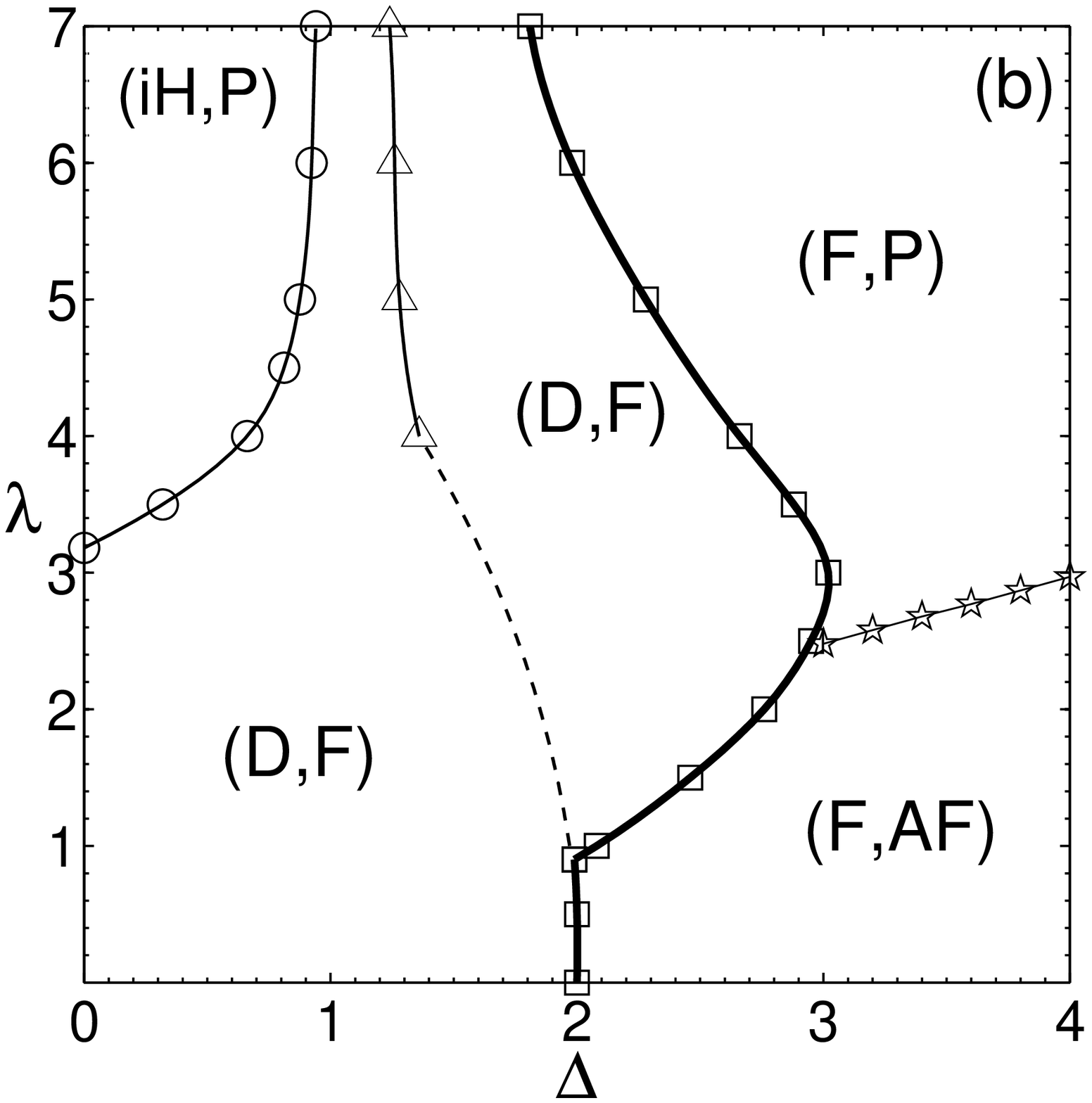}
\caption{Ground state phase diagram of the model~(\ref{Themodel}) for (a)
  $\Delta=0$ and (b) $\alpha=1$. First and second phases in parentheses 
refer to spin  and
orbital sectors, respectively.   
See text, for a description of phases and phase transitions.}
\label{zeroHund}
\end{figure}

{\it Ground-state phase diagram for $\Delta=0$}.-- 
We first discuss the case $\Delta=0$ relevant for realization $\mathsf{A}$.  
At $\lambda= 0$ the orbitals become classical and the ground-state phase 
diagram is easily mapped. For $\alpha>2$, the ground-state is $2\times 2^N$ 
degenerate: there is anti-ferro~(AF) order in orbitals 
$\langle \sigma_i^z \rangle=\pm (-1)^i$ ($\pm$ refers
to two degenerate AF states), 
whereas the spin part is completely 
degenerate (bold line on Fig.~\ref{zeroHund}(a)). For $\alpha > 2$ and 
${\lambda\to 0}$, the system is described by an effective spin model, 
$H_S \sim {\lambda^2}\sum_i  {\bf S}_i{\bf S}_{i+1}$, 
an isotropic Heisenberg antiferromagnet~(iH). Thus, quantum fluctuations of orbitals, induced by $\lambda$-term, lifts immediately the infinite degeneracy of the ground state, resembling order from disorder.
We denote this phase (iH,AF), where the first term denotes iH spin phase, 
and the second term AF orbital phase. We employ 
a similar notation  from now on.
Higher order terms in $\lambda/\alpha$ cannot break the $SU(2)$ spin and
translational symmetries, and thus for $\lambda\to 0$ the iH phase in spin
degrees is stable. 
As shown below, the iH is recovered for strong $\lambda$ independently of 
the value of $\alpha$. Thus, we can expect that for $\alpha>2$ there is an 
unique iH phase in spin degrees of freedom for any $\lambda\neq 0$. On the 
contrary, with increasing $\lambda$ the orbital degrees of freedom experience 
an Ising transition from AF to paramagnetic~(P) phase with 
$\langle \sigma_i^z \rangle=0$. 

For $\lambda=0$ and $0<\alpha<2$, the exact ground state is two-fold
degenerate and represents a direct product of ferro~(F) orbital order, 
$\langle \sigma_{i}^{z} \rangle=+1~(-1)$, 
and spontaneously dimerized Majumdar-Ghosh (MG)
state~\cite{MG} in spins, with spin-singlets located on odd (even) bonds.  
We call this phase a dimer-ferro (D,F). An infinitesimal $\lambda$ generates 
an exchange between the disconnected nearest-neighbour dimers,
$\sim \lambda^4 \sum {\bf S}_{2i+1}{\bf S}_{2i+2}$. With increasing $\lambda$, 
the dimerization order in spins, 
$D=\frac{1}{N}\sum_i|\langle  {\bf S}_{i}{\bf S}_{i+1} - 
{\bf S}_{i}{\bf S}_{i-1}\rangle|$, disappears together with the orbital 
ferro order at a Kosterlitz-Thouless~(KT) phase transition~\cite{KT}.
A numerical ground-state phase diagram of the model~(\ref{Themodel}) for
$\Delta=0$ is presented in Fig.~\ref{zeroHund}(a) 
(details of the numerical simulations are discussed later).
%%%%%%%%%%%%%%%%%%%%%%%%%%%%%
%%%%%%%%%%%%%%%%%%%%%%%%%%

{\it Ground-state phase diagram for $\Delta \neq 0$}.-- 
We now turn to the case of finite  $\Delta$. We focus on the regime 
$\alpha=1$ relevant to realistic condensed-matter 
systems~\cite{Jackeli07,Konstantinovic04,Hikihara,Jackeli09}. 
For $\lambda=0$, a simple calculation shows that the ground state is (F,AF) 
for $\Delta>2(2-\alpha)$ and  (D,F) for $\Delta<2(2-\alpha)$. 
Note, that once the spin dimerization pattern
is spontaneously chosen, the direction of orbitals becomes unambiguously
selected, thus Ising $Z_2$ orbital order is 'slaved' by translation 
symmetry breaking. Whereas for $\Delta=0$ an infinitesimal $\lambda$ induces  
AF spin-exchange $\propto\lambda^4$ on inter-dimer~(weaker) bonds
in the (D,F) phase, for finite $\Delta$ the leading spin-exchange along the
weaker bonds is instead ferromagnetic 
$\sim -\lambda^2 \Delta \sum {\bf S}_{2i+1}{\bf S}_{2i+2}$. Hence there is 
a competition between the $\Delta$ and $\lambda$ terms promoting, 
respectively, a F and AF character of the weaker bonds. 
The character of spin correlations on weak bonds 
$\langle{\bf S}_{2i+1}{\bf S}_{2i+2}\rangle$ 
changes from AF~(small $\Delta$ region) to F 
(larger $\Delta$ region) 
across $\Delta_{MG}= \lambda^2/4+O(\lambda^4)$ line, 
resembling the behaviour across the MG point in the 
spin-$\frac{1}{2}$ $j_1-j_2$ model~\cite{WhiteAffleck}.
 
For $\lambda\rightarrow\infty$ the orbital degrees of freedom are quenched.  
The system becomes equivalent to a $SU(2)$-symmetric spin-$\frac{1}{2}$ chain, 
which up to ${\cal O}(\lambda^{-2})$ is described by the effective Hamiltonian,
\begin{eqnarray}
\label{effectivespinmodel}
  H_S = \sum_{i,n} \left[ j_{n} {\bf S}_i{\bf S}_{i+n}+ 
\Omega ({\bf S}_{i-1}{\bf S}_{i})({\bf S}_{i+1}{\bf S}_{i+2}) \right]
\end{eqnarray}
where $j_1= 2-2\Delta +O(\lambda^{-1})$, $j_2= \lambda^{-1}$, $j_3,
\Omega$ are both of order $\lambda^{-2}$, and the longer range exchanges 
are suppressed as $j_n\sim O(\lambda ^{1-n})$ \cite{supplementary}. 
There are two clear phases in this regime, (iH,P) for 
$1-\Delta\gg \lambda^{-1}$ and (F,P) for $1-\Delta\ll -\lambda^{-1}$. 

One may suspect for large 
%$\lambda\to \infty$ 
$\lambda$ a direct (iH,P) to (F,P) transition with growing $\Delta$. 
This, however, is not the case, as can be
shown by a bosonization analysis of the effective spin
model~\eqref{effectivespinmodel}. Starting from (iH,P) state, 
with increasing $\Delta$ the system necessarily enters first 
into a dimerized state via a KT phase transition at
$\Delta_{KT}\simeq 1-1/2\lambda$. 
At larger $\Delta$, a MG state will be
approximated at $\Delta_{MG}\simeq 1+1/2 \lambda$,
where dimerization will reach the value $D \simeq
3/4$. Increasing further $\Delta$, at $\Delta'\simeq 1+3/2\lambda$,
nearest-neighbor coupling  in Eq.~\eqref{effectivespinmodel} vanishes, 
$j_1(\Delta')=0$, and the $1/\lambda^2$ terms, $j_3$ and $\Omega$, are the
leading ones that couple two sub-chains~\cite{supplementary}. 
Bosonization shows~\cite{supplementary} that, inspite of these terms, 
in the large $\lambda$ limit, the system behaves 
at low energies as two decoupled spin-$\frac{1}{2}$ chains, albeit at 
$\Delta^* =\Delta'+O( \lambda^{-2})$. Hence around the 
$\Delta=\Delta^*$ line the effective spin model in strong coupling is 
described by a $j_1-j_2$ model, where $j_1$ changes sign from antiferro 
(for $\Delta< \Delta^*$) to ferro (for $\Delta> \Delta^*$), whereas $j_2$  
stays positive. Bosonization, supported by recent numerical studies, 
predicts that the ground state of two weakly coupled spin-$\frac{1}{2}$ 
chains is dimerized irrespective of the sign of $j_1$ 
coupling~\cite{Nersesyan,Furukawa,commentSato}.  Hence, in our case there is 
a special fine-tuning line bisecting the dimerized phase, 
$\Delta=\Delta^*$ line, described by double KT phase 
transition~\cite{comment2} where spin dimerization and ferro orbital order 
both vanish. Finally, there is a first order phase transition line separating 
(D,F) and (F,P) states at $\Delta_F\simeq 1+7/2\lambda$.

This sequence of phases and phase transition curves ($\Delta_{KT},
\Delta^*,\Delta_F $), which has been established analytically for large $\lambda$ with the help of effective spin model \eqref{effectivespinmodel}, has been confirmed for  
$\lambda\gtrsim 4$ by numerical simulations (discussed below) of
the original model~\eqref{Themodel}.

{\it Numerical procedures}.-- The phase diagrams depicted in 
Figs.~\ref{zeroHund}(a,b) were obtained 
by means of a combination of Lanczos exact diagonalization, density matrix 
renormalization group simulations based on matrix product states 
(MPS)~\cite{Verstraete} and the infinite time-evolving block decimation 
(iTEBD) algorithm~\cite{Vidal}, confirming the analytical predictions for 
$\lambda \to 0$ and $\lambda\to \infty$.

The KT transition between (D,F) and (iH,P) was extracted by Lanczos method from 
the extrapolation of the level crossing~\cite{Nomura} between the first 
excited singlet of the (D,F) phase and the first excited triplet of the 
(iH,P) phase for systems of up to $N=12$ rungs/wells. Ising transition 
lines were obtained, by MPS simulations, from the peak in the fidelity
susceptibility~\cite{You07}, and that on Fig.~\ref{zeroHund}(b) between (F,AF) and 
(F,P) accurately follows the analytical line $\lambda=\Delta/2+\alpha$.
The first order phase transition into the (F,AF) or (F,P) states 
[bold line in Fig.~\ref{zeroHund}(b)] was 
obtained by MPS method, from the jump of the ground-state total spin 
from $0$ (singlet state) to a fully polarized $N/2$ 
(ferro state)~\cite{addphases}. 
In MPS simulations we have used periodic boundary conditions and system sizes
of up to $N=32$ rungs/wells were considered. For $\lambda\gtrsim 4$, 
the double KT phase transition line bisecting the (D,F) phase 
[see Fig.~\ref{zeroHund}(b)] was determined from
vanishing dimer order using iTEBD method.
For smaller values of $\lambda$ ($1< \lambda \lesssim 4$), 
around the dashed line of Fig.~\ref{zeroHund}(b), 
our simulations indicate an intermediate quadrumerized phase \cite{iTEBD}.

{\it Final remarks}.--
We have assumed above a unit occupation per rung/well, 
for which dipolar interactions were necessary for 
realizing Mott insulating state.  The case of two fermions per rung/well
does not  require dipolar interactions. In the case of the snake-like lattice 
[see Fig.~\ref{lattices}(a)] with two fermions per rung, 
a dimerized state along the rungs (for $\lambda>t$) will be separated 
from a dimerized state along the bonds ($\lambda<t$) by a KT phase 
transition at $\lambda=t$. If in the zig-zag geometry of 
Fig.~\ref{lattices}(b) we place two fermions per well, 
both in $p$ orbitals, then for $\lambda\ll J_H$ due to the Hund's coupling 
a total $S=1$ state will be formed in each well, and the Haldane phase of 
a spin-$1$ chain will be realized.

In conclusion, dipolar fermions on zig-zag lattices can capture relevant 
spin-orbital models of realistic $d$-electron systems and allow to 
explore parameter regimes which are hardly accessible for solid-state 
compounds. Moreover, the quantum nature of orbital fluctuations 
can be revived, which combined with geometric frustration and spin dynamics 
produces an intriguing rich ground state phase diagram.

We thank A. Cojuhovschi, G. Khaliullin, A. Kolezhuk, and S. Ospelkaus 
for useful discussions, and S. Furukawa for correspondence on the 
Haldane-Dimer state. This work has been
supported by QUEST (Center for Quantum Engineering and 
Space-Time Research). Support from GNSF/ST09-447 (G.J.) and 
SCOPES Grant IZ73Z0-128058 (T.V.) is acknowledged.
%%%%%%%%%%%%%%

\section{ Supplementary material to ''Exploring spin-orbital models
with dipolar fermions in zig-zag optical lattices'' }

In this supplementary material, we provide additional details 
concerning the derivation of the spin-orbital Hamiltonian and 
the bosonization procedure.

\section{The effective spin-orbital Hamiltonians}
In this section we outline the essential steps of the derivation of the 
effective spin-orbital Hamiltonian for the two systems, 
$\mathsf{A}$ and $\mathsf{B}$, discussed in the Letter. 

\subsection{$\mathsf{A}$:  Snake-like lattice}
We introduce the fermionic annihilation operator $c_{a\,i,s}$, where $a=1~(2)$
indicates the up~(down) sites on $i$-th rung, and $s=\uparrow$ or $\downarrow$ 
refer to the spin. Using the notation of the Letter, system $\mathsf{A}$ 
is described  by the Hubbard like Hamiltonian $H=H_{kin}+H_{int}$, where

\begin{eqnarray}
H_{kin}&=&-\lambda \sum_{i,s} (c^{\dagger}_{1\, i,s} c_{2
  \,i,s}+c^{\dagger}_{2 \,i,s}  c_{1\, i,s}  ) \label{eq1s}\\
&-&\frac{t}{2}\sum_{i,a,s}\bigl[1+(-1)^{i+a}\bigr]\bigl[c^{\dagger}_{a\,i,s}
c_{a\,i+1,s}+ c^{\dagger}_{a\,i+1,s}   c_{a\,i,s}\bigr], \nonumber
\end{eqnarray}
accounts for single-particle processes, and
\begin{eqnarray}
H_{int}&=&U\sum_{i,a}  c^{\dagger}_{a\, i,\uparrow}  c_{a\,
  i,\uparrow}c^{\dagger}_{a\,i,\downarrow} c_{a\,i,\downarrow}\label{eq2s}\\
&+&V \sum_{i,s,s^{\prime}} c^{\dagger}_{1\,i,s}  
c_{1\,i ,s}c^{\dagger}_{2\,i,s^{\prime}} c_{2\,i ,s^{\prime}}.\nonumber 
\end{eqnarray}
corresponds to two-particle interactions. In the limit $U,V \gg t$, and 
retaining one particle per ladder rung (quarter filling), 
the effective Hamiltonian up to the second order in $t$ acquires 
the form of Eq.~(1) of the Letter~(at $\Delta=0$), where 
\begin{eqnarray}
S^z_i&=&\sum_{a}S^z_{a\, i}= 
\frac{1}{2}\sum_{a}(c_{a\,i,\uparrow}^{\dagger}c_{a\, i,\uparrow}-
c^{\dagger}_{a\, i\downarrow}c_{a\, i,\downarrow}) \nonumber\\
S^+_i&=& \sum_{a} S^+_{a \,i} = \sum_{a}c^{\dagger}_{a\, i,\uparrow}c_{a\,
  i,\downarrow}~, 
\nonumber
\end{eqnarray}
are the spin-$\frac{1}{2}$ operators on the $i$-th rung, and 
\begin{eqnarray}
\sigma^x_i&=&\sum_{s}(c^{\dagger}_{1\, i,s}c_{2\, i,s}+c^{\dagger}_{2\,
  i,s}c_{1\, i, s}) \nonumber\\
\sigma^z_i&=&\sum_{s}(c^{\dagger}_{1\, i,s}c_{1\, i,s}-c^{\dagger}_{2\,
  i,s}c_{2\, i,s})
\nonumber
\end{eqnarray}
are the Pauli matrices corresponding to the pseudo-orbital degrees of
freedom. 

\subsection{$\mathsf{B}$:  $p$-band zig-zag lattice}
In the system $\mathsf{B}$, the pseudo orbital index  $a=1~(2)$ refers  to $p_x$
($p_y$) orbital in a given well. The kinetic part of the Hamiltonian  remains
the same as in Eq.~(\ref{eq1s}), while the interaction part, in addition to the
terms shown in Eq.~(\ref{eq2s}), acquires the following supplementary terms:
\begin{eqnarray}
H_{int}^{\prime}&=&-2J_H\sum_{i}\left[{\bf S}_{1\,i}{\bf
  S}_{2\,i}
+\frac{n_{1\,i}n_{2\,i}}{4}\right]\label{eq3s}\\
&+&J_{\rm H}\sum_{i}[
c_{1\,i\uparrow}^{\dagger}c_{1\,i\downarrow}^{\dagger}
c_{2\,i\downarrow}c_{2\,i\uparrow}+c_{2\,i\uparrow}^{\dagger}c_{2\,i\downarrow}^{\dagger}
c_{1\,i\downarrow}c_{1\,i\uparrow}],\nonumber
\end{eqnarray}
where ${\bf S}_{1(2)\,i}$ and $n_{1(2)\,i}$ are, respectively, spin and
density operators for a fermion in  $p_x$($p_y$) orbital state.
The first line stands for Hund's exchange and the second one describes 
the so-called pair-hopping~\cite{CNR}.

The coupling constants $U$, $V$, and $J_H$, entering in 
Eqs.~(\ref{eq2s},\ref{eq3s}),  can be expressed in terms of a 
two-body potential and single-particle wave functions~\cite{CNR}. One finds:
\begin{eqnarray}
&&U\!=\!\!\int\mathrm{d} \mathbf {r_1} \mathrm{d} \mathbf {r_2}
  {P^2_{x(y)}}(\mathbf {r_1})V(\mathbf {r_1}- \mathbf {r_2}) {P^2_{x(y)}}(\mathbf {r_2}),\label{eq:U}\\
&&V\!=\!\!\int\mathrm{d} \mathbf {r_1} \mathrm{d} \mathbf {r_2}
 {P^2_x}(\mathbf {r_1})V(\mathbf {r_1}- \mathbf {r_2})
 {P^2_y}(\mathbf {r_2}),
\label{eq:V}
\end{eqnarray}
and
\begin{eqnarray}
J_H\!=\!\!\int\!\!\mathrm{d} \mathbf {r_1} \mathrm{d} 
\mathbf {r_2}{P_x}(\mathbf{r_1})  {P_y}(\mathbf {r_1}) 
V(\mathbf {r_1}- \mathbf {r_2}){P_x}(\mathbf{r_2}) {P_y}(\mathbf {r_2}).
\label{eq:JH}
\end{eqnarray} 
Above $P_x$ and $P_y$ are orbital wavefunctions of the same well, and
$V(\mathbf {r_1}- \mathbf {r_2})$ is the total interparticle potential
(including both contact as well as dipolar interactions).

The energy of the state with  two fermions occupying the same orbital is $U$.
However, the pair-hopping term moves two particles from the doubly occupied 
orbital to an empty orbital with the amplitude $J_H$, and true eigenstates 
become  orbital-symmetric and orbital-antisymmetric states with 
corresponding energies $U+J_H$ and $U-J_H$.

Two fermions occupying different orbitals may form a spin-singlet or 
a spin-triplet state with corresponding energies 
$V+J_H$ and $V-J_H$~\cite{comment}.

In the limit $U\pm J_H,V\pm J_H\gg t$, and up to second order in $t$, 
the effective Hamiltonian acquires the form
$H=\sum_i H_{i,i+1}$, where $H_{i,i+1}$ is the two-site Hamiltonian:
\begin{eqnarray}
\label{suppThemodel}
H_{i,i+1}&=\!& \!\!-\frac{t^2}{\tilde U}P_{i,i+1}(S^T\!\!\!=\!0)(1\!+\!(-1)^i 
\sigma_i^z)(1\!+\!(-1)^i \sigma_{i+1}^z)\nonumber\\
&-& \frac{t^2}{2(V+J_H)}P_{i,i+1}(S^T\!\!=\!0) (1- \sigma_i^z \sigma_{i+1}^z)  
\nonumber\\
&-&  \frac{t^2}{2(V-J_H)}P_{i,i+1}(S^T\!\!=\!1)  (1- \sigma_i^z
\sigma_{i+1}^z)  
\nonumber\\
&-&\frac{\lambda}{2} (\sigma_i^x+ \sigma_{i+1}^x ),
\end{eqnarray}
where we have introduced the operators 
$P_{i,i+1}(S^T\!\!=\!0)=- {\bf S}_i{\bf S}_{i+1} +1/4 $, and 
$P_{i,i+1}(S^T\!\!=\!1)={\bf S}_i{\bf S}_{i+1} +3/4 $, 
which project onto two-fermion states on sites $i$ and $i+1$ with, 
respectively, total spin $S^T=0$ and $S^T=1$, and ${\tilde U}=(U^2-J_{H}^{2})/U$.

The first line in Eq.~\eqref{suppThemodel} accounts for the situation in 
which both fermions occupy same orbitals on sites $i$ and $i+1$. In that case, 
when a fermion hops to the neighboring site~\cite{hoppingcomment}, 
the orbital-symmetric or orbital-antisymmetric states of the spin-singlet pair
 is reached (note that the formation of the spin-triplet pair is forbidden by 
Pauli exclusion).

The second and third lines of Eq.~\eqref{suppThemodel} account for the
configuration in which two fermions occupy orthogonal orbitals on neighboring
sites. When a fermion hops to a (singly) occupied neighboring site
(remembering that inter-site hopping $t$ does not change orbital quantum
number), depending on the total spin of the two fermions the interaction may 
take place either in the spin-singlet channel, with energy cost $V+J_H$, 
or in the spin-triplet channel, with energy cost $V-J_H$. A proper re-ordering 
of the terms in Eq.~\eqref{suppThemodel} leads to the Hamiltonian of Eq.~(1) 
of the Letter.

\section{Bosonization analysis}
In the following we provide some details on our bosonization analysis of 
the large $\lambda$ regime ($\lambda\rightarrow\infty$), and in particular 
on how this approach provides an understanding of the special fine-tuning 
line $\Delta=\Delta^*$ cutting the (D,F) phase at which spin dimerization 
and orbital ferro orders vanish.

%%%%%%%%%%%%%%%%
\begin{figure}%[ht]
\includegraphics[width=7.6cm]{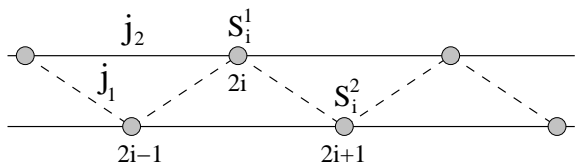}
\caption{Two sub-chains with intrachain coupling $j_2$ (bold lines) coupled 
by zig-zag interchain coupling $j_1$ (dashed line).}
\label{latticesS}
\end{figure}
%%%%%%%%%%%%%%%%%, 
Integrating out the orbitals in large  $\lambda$ limit, 
to the second order in $1/\lambda$ we obtain the effective spin-$\frac{1}{2}$ 
model given in Eq. (2) of the Letter with,
\begin{eqnarray}
j_1&=& 2(1-\Delta)+ \frac{4+\Delta(1+\Delta)}{2\lambda }+O(\lambda^{-2}),
\nonumber\\
j_2&=&\frac{1}{\lambda}+O(\lambda^{-2}), \,\,\, j_3= 
\frac{1+\Delta}{2\lambda ^2}+O(\lambda^{-3}),\nonumber\\
\Omega&=&  \frac{2}{\lambda ^2}+O(\lambda^{-3})~,
\end{eqnarray}
where we use the same notations as in the Letter.
$\lambda\to \infty$ is a convenient limit for bosonization, 
in particular for $j_1\ll j_2$. In that case, one can consider two weakly 
coupled chains~(see Fig.~\ref{latticesS}), which we denote as 
${\bf S}_{i}^1={\bf S}_{2i}$ and ${\bf S}_{i}^2={\bf S}_{2i+1}$.
Using bosonization identification, the spins along each chain are 
represented by smooth and staggered parts
$
 {\bf S}_{i}^a\to {\bf J}_L^a+{\bf J}_R^a+(-1)^i{\bf n}^a
$~\cite{GNT}.
The coupling of the two chains induces marginal operators. Apart from the 
scalar (dimerization) operators, there are spinfull marginal terms, 
so-called twist operators~\cite{Nersesyan98}. Up to irrelevant and marginal 
couplings that induce only velocity renormalization, the effective 
bosonized Hamiltonian density describing the spin degrees of freedom 
in the large-$\lambda$ limit is of the form
\begin{eqnarray}
\label{EffectiveBosonization}
&&{\cal H}= \frac{\pi v}{2}{\bf J}_\xi^a{\bf J}_\xi^a+ 
D_1({\bf J}_L^1{\bf J}_R^2+{\bf J}_R^1{\bf J}_L^2 )+ D_2 {\bf J}_L^a{\bf J}_R^a \\
&&+T_{1}\varepsilon_{ab} (\epsilon^a\partial_x{ \epsilon}^b  + 
{\bs n}^a\partial_x{\bs n}^b)+T_{2}\varepsilon_{ab} 
(3 \epsilon^a\partial_x{ \epsilon}^b  - {\bs n}^a\partial_x{\bs n}^b),\nonumber
\end{eqnarray}
where $v \sim j_2$, a summation convention is assumed for repeated indices 
$a,b=(1,2)$, and $\xi=(L,R)$, $\varepsilon_{ab}$ is the antisymmetric symbol, 
and $\epsilon^a$ stand for the dimerization operators, such that 
${\bf S}_{i}^a {\bf S}_{i+1}^a\sim (-1)^i{ \epsilon}^a(x)+$ a less relevant 
smooth part. 

Bare values of the twist couplings $T_{1,2}$, and the 
inter-chain dimerization $D_1$ depend on $j_1,j_3$ and $\Omega$, in 
linear order (with non-universal proportionality coefficients), whereas 
the intra-chain dimerization amplitude $D_2$ depends on $j_2$ and $\Omega$.

One-loop RG equations corresponding to the effective model 
\eqref{EffectiveBosonization} are identical to those obtained 
in~[\onlinecite{Nersesyan98}] for the $SU(2)$ symmetric case. RG flow is 
dominated by interchain dimerization $D_1$, which in the infrared limit 
always scales to strong coupling, except for the initial condition $T_1=T_2$, 
corresponding to the fine tuning $\Delta= \Delta^*$, which we interpret as 
an infrared decoupling point of the two chains. In contrast $D_2$ scales to 
zero at low energies for $\lambda\to \infty$. Thus for $\lambda\to \infty$ 
there is a special line in (D,F) phase where spin dimerization 
(and consequently the ferro orbital order) vanishes.

In fact, bosonization is only needed to capture the influence of $j_3$ and 
$\Omega$ terms close to $j_1=0$. Away of that region, for $\lambda\to \infty$, 
one just needs to retain $j_1$ and $j_2$ terms, and borrow known results 
from the frustrated $j_1-j_2$ spin-$\frac{1}{2}$ chain.
This gives us an estimate of other phase transition lines  
(we use the notations of the Letter) $\Delta_{KT}$ and $\Delta_F$ which
follow respectively from $j_1\simeq 4j_2$ and $j_1=-4j_2$ conditions. 
Whereas $\Delta_{MG}$ (which does not represent a phase transition line) 
follows from the condition $j_1=2j_2$.


\begin{thebibliography}{10}
\bibitem{footnote-lattices} Ladder-like (zig-zag) lattices may be 
obtained by combining two-well potentials and square (triangular) lattices.
\bibitem{addphases} For $\lambda> 1 $,  before reaching ferro spin state 
we could not resolve any more dimer order of (D,F) state. We thus cannot 
reliably rule out the existence of an additional phase 
(the ground state though remains 
a global spin singlet)  in a narrow region before 
reaching the  fully polarized spin state (F,AF) or (F,Para).
\bibitem{iTEBD} As indicated by our iTEBD simulations, for 
$1<\lambda  \lesssim 4$ (around the dashed line of Fig.2(b)), 
there is an intermediate 4-fold degenerate phase, 
with a unit cell composed of 4 sites, where apart of the (spontaneous) 
quadrumerized pattern of nearest neighbour spin correlations, 
next-nearest spin correlations have a finite dimer order, and 
orbitals show up-up-down-down ferrimagnetic pattern, with 
finite $\sum_i\langle \sigma_i^z\rangle\neq 0$. It however demands considerable 
numerical resourses to determine accurately the boundaries of this phase, 
that we leave to future large-scale numerical investigations.
\end{thebibliography}
\begin{thebibliography}{10}
\bibitem{comment} For the spin-triplet state, the coordinate wave-function must 
have a node when the positions of the two fermions coincide. Hence, for the 
spin-triplet state the contact $s$-wave interaction does not play any role. 
This is reflected in Eqs.~\eqref{eq:V} and \eqref{eq:JH} where for 
$V(\mathbf {r_1}- \mathbf{r_2})\sim \delta(\mathbf {r_1}- \mathbf{r_2})$,  
$V=J_H$ and hence scattering in the spin-triplet channel vanishes.
\bibitem{hoppingcomment} Note that hopping from a site $i$ to a site $i+1$ is 
only possible if the fermion occupies on site $i$ the orbital that is 
directed towards the site $i+1$, as depicted on Fig. 1(b) of the Letter.
\end{thebibliography}
\end{document}